\documentclass{article}

\usepackage{graphics}
\usepackage{graphicx}
\usepackage{epsfig}

\newcommand{\conf}[2]{\ensuremath{(A, #1; B, #2)}}
\newcommand{\T}[1]{\ensuremath{\bar{#1}}}
\newcommand{\emptytuple}{\ensuremath{\langle\rangle}}
\newcommand{\semijoin}{\ensuremath{\ltimes}}
\newcommand{\ARITY}{\ensuremath{\mathrm{arity}}}
\newcommand{\game}[3]{\ensuremath{G_{#1}\conf{#2}{#3}}}
\newcommand{\dbschema}{\ensuremath{\mathbf{S}}}

\renewcommand{\leq}{\leqslant}
\renewcommand{\geq}{\geqslant}

\usepackage{amsmath}
\usepackage{amssymb}
\usepackage{amsthm}

\theoremstyle{plain} 
\newtheorem{theorem}{Theorem}
\newtheorem{lemma}[theorem]{Lemma}
\newtheorem{proposition}[theorem]{Proposition}
\newtheorem{corollary}[theorem]{Corollary}

\theoremstyle{definition}
\newtheorem{definition}[theorem]{Definition}

\theoremstyle{remark}

\begin{document}

\title{The semijoin algebra and the guarded fragment}
\date{} 
\author{Dirk
  Leinders\\ Jerzy Tyszkiewicz~\footnote{This author has been
    partially supported by the European Community Research Training
    Network ``Games and Automata for Synthesis and Validation''
    (GAMES), contract HPRN-CT-2002-00283.}\\ Jan Van den Bussche}
\maketitle

\begin{abstract}
  The semijoin algebra is the variant of the relational algebra
  obtained by replacing the join operator by the semijoin operator. We
  discuss some interesting connections between the semijoin algebra
  and the guarded fragment of first-order logic. We also provide an
  Ehrenfeucht-Fra\"{\i}ss\'{e} game, characterizing the discerning
  power of the semijoin algebra. This game gives a method for showing
  that certain queries are not expressible in the semijoin algebra.
\end{abstract}

\section{Introduction}
\label{sec:introduction}

Semijoins are very important in the field of database query
processing. While computing project-join queries in general is
NP-complete in the size of the query and the database, this can be
done in polynomial time when the database schema is
acyclic~\cite{Yannakakis81}, a property known to be equivalent to the
existence of a semijoin program~\cite{BFMY83}. Semijoins are often
used as part of a query pre-processing phase where dangling tuples are
eliminated.  Another interesting property is that the size of a
relation resulting from a semijoin is always linear in the size of the
input. Therefore, a query processor will try to use semijoins as often
as possible when generating a query plan for a given query (a
technique known as ``pushing projections''~\cite{GMUW00}). Also in
distributed query processing, semijoins have great importance, because
when a database is distributed across several sites, they can help
avoid the shipment of many unneeded tuples.

Because of its practical importance, we would like to have a clear
knowledge of the capabilities and the limitations of semijoins. For
example, Bernstein, Chiu and Goodman~\cite{BC81,BG81} have
characterized the conjunctive queries computable by semijoin programs.
In this paper, we consider the much larger class of queries computable
in the variant of the relational algebra obtained by replacing the
join operator by the semijoin operator. We call this the semijoin
algebra (SA). A join of two relations combines all tuples satisfying a
given condition, called the join condition. A semijoin differs from a
join in the sense that it selects only those tuples in the first
relation that participate in the join. The semijoin algebra is a
fragment of the relational algebra, which is known to be equivalent to
first-order logic (called relational calculus in database
theory~\cite{AHV95}).

Interestingly, there is a fragment of first-order logic very similar
to the semijoin algebra: it is the so called ``guarded fragment''
(GF)~\cite{AvBN98,Graedel99,GHO00,Hirsch02}, which has been studied in
the field of modal logic. This is interesting because the motivations
to study this fragment came purely from the field of logic and had
nothing to do with database query processing.  Indeed, the purpose was
to extend propositional modal logic to the predicate level, retaining
the good properties of modal logic, such as the finite model property.
An important tool in the study of the expressive power of the GF is
the notion of ``guarded bisimulation'', which provides a
characterization of the discerning power of the GF\@.

We will show that when we allow only equalities to appear in the
semijoin conditions, the semijoin algebra has essentially the same
expressive power as the guarded fragment. When also nonequalities or
other predicates are allowed, the semijoin algebra becomes more
powerful. We will define a generalization of guarded bisimulation, in
the form of an Ehrenfeucht-Fra\"{\i}ss\'{e} game, that characterizes the
discerning power of the semijoin algebra. We will use this tool to
show that certain queries are not expressible in SA.

\section{Preliminaries}
\label{sec:preliminaries}

In this section, we give formal definitions of the semijoin algebra
and the guarded fragment.

From the outset, we assume a universe $\mathbb{U}$ of basic data
values, over which a number of predicates are defined. These
predicates can be combined into quantifier-free first-order formulas,
which are used in selection and semijoin conditions. The names of
these predicates and their arities are collected in the vocabulary
$\Omega$.  The equality predicate ($=$) is always in $\Omega$. A
database schema is a finite set $\dbschema$ of relation names, each
associated with its arity.  $\dbschema$ is disjoint from $\Omega.$ A
database $D$ over $\dbschema$ is an assignment of a finite relation
$D(R) \subseteq \mathbb{U}^n$ to each $R \in \dbschema$, where $n$ is
the arity of $R$.

\textbf{Proviso.} When $\varphi$ stands for a first-order formula, then
$\varphi(x_1,\ldots,x_k)$ indicates that all free variables of $\varphi$ are
among $x_1,\ldots,x_k$.

First, we define the Semijoin Algebra.

\begin{definition}[Semijoin algebra, SA] Let $\dbschema$ be a database
  schema. Syntax and semantics of the Semijoin Algebra is inductively
  defined as follows:
  \begin{enumerate}
  \item Each relation $R \in \dbschema$ is a semijoin algebra expression.
  \item If $E_1, E_2 \in $ SA have arity $n$, then also $E_1 \cup
    E_2$, $E_1 - E_2$ belong to SA and are of arity $n$.
  \item If $E\in $ SA has arity $n$ and $X \subseteq \{ 1,\ldots,n\}$,
    then $\pi_X(E)$ belongs to SA and is of arity $\#X$.
  \item If $E_1, E_2 \in$ SA have arities $n$ and $m$, respectively, and
    $\theta_1(x_1,\ldots,x_n)$ and
    $\theta_2(x_1,\ldots,x_n,y_1,\ldots,y_m)$ are quantifier-free
    formulas over $\Omega$, then also $\sigma_{\theta_1} (E_1)$ and
    $E_1 \semijoin_{\theta_2} E_2$ belong to SA and are of arity $n.$
  \end{enumerate}
  
  The semantics of the projection, the selection and the semijoin
  operator are as follows: $\pi_X(E) := \{ (a_i)_{i \in X} \mid
  (a_1,\ldots,a_n) \in E \}$, $\sigma_{\theta_1} (E) :=
  \{(a_1,\ldots,a_n)\in E \mid \theta_1(a_1,\ldots, a_n)$ holds$\}$,
  $E_1 \semijoin_{\theta_2} E_2 := \{ (a_1,\ldots,a_n) \in E_1 \mid
  \exists (b_1,\ldots, b_m) \in E_2$,
  $\theta_2(a_1,\ldots,a_n,b_1,\ldots,b_m)$ holds$\}$. The semantics
  of the other operators are well known.
\end{definition}

Now, we recall the definition of the guarded fragment.

\begin{definition}[Guarded fragment, GF]
  Let $\dbschema$ be a database schema.
  \begin{enumerate}
  \item All quantifier-free first-order formulas over
    $\dbschema$ are formulas of GF.
  \item If $\varphi$ and $\psi$ are formulas of GF, then so are $\lnot
    \varphi$, $\varphi \lor \psi$, $\varphi \land \psi$, $\varphi
    \rightarrow \psi$ and $\varphi \leftrightarrow \psi$.
  \item If $\varphi(\T{x},\T{y})$ is a formula of GF and
    $\alpha(\T{x},\T{y})$ is an atomic formula such that all
    free variables of $\varphi$ do actually occur in $\alpha$ then
    $\exists \T{y} (\alpha(\T{x},\T{y}) \land
    \varphi(\T{x},\T{y}))$ is a formula of GF.
  \end{enumerate}
  As the guarded fragment is a fragment of first-order logic,
  the semantics of GF is that of first-order logic, interpreted
  over the active domain of the database~\cite{AHV95}.
\end{definition}

\section{Semijoin algebra versus guarded fragment}
\label{sec:SAvsGF}

In this section, $\Omega = \{ = \}$ consists only of the equality predicate.
Suppose furthermore that we only allow conjunctions of equalities to
be used in the semijoin conditions; selection conditions can be
arbitrary quantifier-free formulas over $\Omega$. We will denote the
semijoin algebra with this restriction on the semijoin conditions by
SA$^=$. Before we prove that SA$^=$ is subsumed by GF, we need a lemma.

\begin{lemma} \label{lem:SAtuple}
  For every SA$^=$ expression $E$ of arity $k$, for every database $A$
  and for every tuple $\T{z} = (z_1,\ldots , z_k)$ in $E(A)$, there
  exists $R$ in $\mathbf{S}$, an injective function $f:\{1,\ldots,k\}
\rightarrow \{1,\ldots,\ARITY(R) \}$, and a tuple $\T{t}$ in $A(R)$
such that $\bigwedge_{i=1}^{k} z_i = t_{f(i)}$.
\end{lemma}
\begin{proof}
  By structural induction on expression $E$.
\end{proof}

\begin{theorem}\label{SAsubsumesGF}
  For every SA$^=$ expression $E$ of arity $k$, there exists a GF
  formula $\varphi_E$ such that for every database
  $D$, $E(D) = \{ \T{d} \in D \mid \varphi_E(\T{d})\}$.
\end{theorem}
\begin{proof}
  The proof is by structural induction on $E$.
  \begin{itemize}
  \item if $E$ is $R$, then $\varphi_E(x_1,\ldots,x_k) :=
    R(x_1,\ldots,x_k)$.
  \item if $E$ is $E_1 \cup E_2$, then $\varphi_E(x_1,\ldots,x_k) :=
    \varphi_{E_1}(x_1,\ldots,x_k)\ \lor\ 
    \varphi_{E_2}(x_1,\ldots,x_k)$.
  \item if $E$ is $E_1 - E_2$, then $\varphi_E(x_1,\ldots,x_k) :=
    \varphi_{E_1}(x_1,\ldots,x_k)\ \land\ 
    \lnot\varphi_{E_2}(x_1,\ldots,x_k)$.
  \item if $E$ is $\sigma_{\theta}(E_1)$, then
    $\varphi_E(x_1,\ldots,x_k) := \varphi_{E_1}(x_1,\ldots,x_k) \land
    \theta(x_1,\ldots,x_k)$.
  \item if $E$ is $\pi_{i_1,\ldots,i_k}(E_1)$ with $E_1$ of arity $n$,
    then, by induction, $\varphi_{E_1}(z_1,\ldots,z_n)$ defines all
    tuples in $E_1(D)$. By Lemma~\ref{lem:SAtuple},
    $\varphi_{E_1}(\T{z})$ is equivalent to the formula obtained by replacing in $\psi := $
    \[
    \bigvee_{R \in \dbschema} \bigvee_{\substack{f:\{1,\ldots,n\} \rightarrow \\ \{1,\ldots,\ARITY(R) \}}} 
    \exists (t_j)_{j \in Q}
    \big(
    R(\T{t})\ \land\ \varphi_{E_1}(t_{f(1)},\ldots,t_{f(n)})
    \big)
    \]
    each $t_{f(i)}$ by $z_i$, $i=1,\ldots, n$.  In this formula, $Q$
    is a shorthand for the set $\{1,\ldots,\ARITY(R) \} -
    f(\{1,\ldots,n\})$.  Formula $\varphi_E$ should now only select
    components $i_1,\ldots,i_k$ out of this formula. To this end, we
    modify $\psi$ such that in each disjunct it quantifies over
    $(t_j)_{j \in Q'}$ with $Q' = \{1,\ldots,\ARITY(R)\} -
    f(\{i_1,\ldots,i_k \})$ and in each disjunct $t_{f(i_l)}$ is
    replaced by $x_l$, $l=1,\ldots,k$. Now $\varphi_E(x_1,\ldots,x_k)$
    is obtained.
  \item if $E$ is $E_1 \semijoin_{\theta} E_2$ with $\theta =
    \bigwedge_{l=1}^s x_{i_l} = y_{j_l} $ and $E_2$ of arity $n$,
    then, by induction, $\varphi_{E_1}(x_1,\ldots,x_k)$ and
    $\varphi_{E_2}(z_1,\ldots,z_n)$ define all tuples in $E_1(D)$ and
    $E_2(D)$ respectively.  By Lemma~\ref{lem:SAtuple},
    $\varphi_E(x_1,\ldots,x_k)$ is obtained by replacing in formula
    $\chi := $
    \begin{multline*}
    \phi_{E_1}(x_1,\ldots,x_k)\ \land \\
    \bigvee_{R \in \dbschema} \bigvee_{\substack{f:\{1,\ldots,n\} \rightarrow \\ \{1,\ldots,\ARITY(R) \}}} 
    \exists (t_j)_{j \in Q''}
    \big(
    R(\T{t})\ \land\ \varphi_{E_2}(t_{f(1)},\ldots,t_{f(n)})
    \big)
    \end{multline*}
    each $t_{f(j_l)}$ by $x_{i_l}$, $l=1,\ldots,s$. Note that
    condition $\theta$ is enforced by repetition of variables
    $x_{i_l}$. In this formula, $Q'' = \{1,\ldots,\ARITY(R) \} -
    f(\{j_1,\ldots,j_s \})$.
  \end{itemize}
\end{proof}

By the decidability of GF, we obtain:

\begin{corollary}
  Satisfiability of SA$^=$ expressions is decidable.
\end{corollary}

With decidability of SA expressions, we always mean finite
satisfiability, because a database is finite by definition.

The literal converse statement of Theorem~\ref{SAsubsumesGF} is not
true, because the guarded fragment contains all quantifier-free
first-order formulas, so that one can express arbitrary cartesian
products in it, such as $\{ (x,y) \mid S_1(x) \land S_2(y) \}$.
Cartesian products, of course, can not be expressed in the semijoin
algebra. Nevertheless, the result of any GF query restricted to a
single relation by a semijoin is always expressible in SA$^=$:

\begin{theorem}
  For every GF formula $\varphi(x_1,\ldots,x_k)$, for every relation $R$
  (with arity $n$), for every injective function $f:\{1,\ldots,k\}
  \rightarrow \{1,\ldots,n\}$, the query $\{\T{x}\mid \varphi(\T{x}) \}
  \semijoin_{\theta} R$ in which $\theta$ is $\bigwedge_{i=1}^k x_i =
  y_{f(i)}$, is expressible in SA$^=$.
\end{theorem}
\begin{proof}
  By structural induction on $\varphi$, we construct the desired semijoin
  expression $E_{\varphi,k}^{f,R}$.
  \begin{itemize}
  \item if $\varphi(x_1,\ldots,x_k)$ is $T(x_{i_1},\ldots,x_{i_l})$ then
    $E_{\varphi,k}^{f,R}:= \pi_{f(1),\ldots,f(k)}(R) \semijoin_{\theta}
    T$, where $\theta$ is $(x_{i_1} = y_1) \land (x_{i_2} = y_2) \land
    \ldots \land (x_{i_l} = y_l)$;
  \item if $\varphi(x_1,\ldots,x_k)$ is  $(x_i = x_j)$ then
    $E_{\varphi,k}^{f,R}:= \sigma_{i=j}(\pi_{f(1),\ldots,f(k)}(R))$;
  \item if $\varphi(x_1,\ldots,x_k)$ is $\psi(x_1,\ldots,x_k) \lor
    \xi(x_1,\ldots,x_k)$ then $E_{\varphi,k}^{f,R} := E_{\psi,k}^{f,R}
    \cup E_{\xi,k}^{f,R}$;
  \item if $\varphi(x_1,\ldots,x_k)$ is $\lnot \psi(x_1,\ldots,x_k)$ then
    $E_{\varphi,k}^{f,R} := \pi_{f(1),\ldots,f(k)}(R) -
    E_{\psi,k}^{f,R}$;
  \item suppose $\varphi(x_1,\ldots,x_k)$ is  $\exists \T{z}
    (\alpha(\T{x},\T{z}) \land \psi(\T{x},\T{z}))$, where $\alpha$ is
    atomic with relation name $T$.  Let $x_{i_1},\ldots,x_{i_r}$ be the
    different occurrences of variables among $x_1,\ldots,x_k$ in
    $\alpha$. Now, $E_{\varphi,k}^{f,R} := \pi_{f(1),\ldots,f(k)}(R)
    \semijoin_{\theta} E_{\psi,r+l}^{g,T}$ where $\theta$ is $(x_{i_1}
    = y_1) \land (x_{i_2} = y_2) \land \ldots \land (x_{i_r} = y_r)$
    and $g$ is the function that maps $j \in \{1,\ldots,r\}$ to the
    position of $x_{i_j}$ in $\alpha$ and that maps $j \in
    \{r+1,\ldots,r+l\}$ to the position of $z_{j-r}$ in $\alpha$.
\end{itemize}
\end{proof}

Taking $k=0$ and $R$ equal to any nonempty relation in the above
theorem, we obtain:

\begin{corollary}
  Over the class of nonempty databases GF sentences and 0-ary SA$^=$
  expressions have equal expressive power.
\end{corollary}

Here, a database is said to be empty if all its relations are empty.

Let us now allow arbitrary semijoin conditions (still over equality
only). Specifically, nonequalities are now allowed. We will denote the
semijoin algebra over $\Omega = \{ = \}$ by SA$^{\neq}$. Then, GF no
longer subsumes SA$^{\neq}$. A counterexample is the query that asks
whether there are at least two distinct elements in a single unary
relation $S$.  This is expressible in SA$^{\neq}$ as $S \semijoin_{x_1
  \neq y_1} S$, but is not expressible in GF.  Indeed, a set with a
single element is guarded bisimilar to a set with two
elements~\cite{AvBN98,Hirsch02}.

Unfortunately, these nonequalities in semijoin conditions make SA undecidable.

\begin{theorem}
  Satisfiability of SA$^{\neq}$ expressions is undecidable.
\end{theorem}
\begin{proof}
  Gr\"adel~\cite[Theorem 5.8]{Graedel99} shows that GF with
  functionality statements in the form of functional[$D$], saying that
  the binary relation $D$ is the graph of a partial function, is a
  conservative reduction class. Since functional[$D$] is expressible
  in SA$^{\neq}$ as $D \semijoin_{x_1=y_1 \land x_2 \neq y_2} D =
  \emptyset$, it follows that SA$^{\neq}$ is undecidable.
\end{proof}

In the next section, we will generalize guarded bisimulation to the
semijoin algebra, with arbitrary quantifier-free formulas over
$\Omega$ as semijoin conditions.

\section{An Ehrenfeucht-Fra\"{\i}ss\'{e} game for the semijoin algebra}
\label{sec:semijoingame}

In this section, we describe an Ehrenfeucht-Fra\"{\i}ss\'{e} game that
characterizes the discerning power of the semijoin algebra.

Let $A$ and $B$ be two databases over the same schema $\dbschema$. The
\emph{semijoin game} on these databases is played by two players,
called the spoiler and the duplicator. They, in turn, choose tuples
from the tuple spaces $T_A$ and $T_B$, which are defined as follows:
$T_A := \bigcup_{R \in \dbschema} \bigcup \big\{ \pi_{\mathrm{X}}(A(R))
  \mid \mathrm{X} \subseteq \{1,\ldots,\ARITY(R)\} \big\}$, and $T_B$
is defined analogously. So, the players can pick tuples from the
databases and projections of these.
 
At each stage in the game, there is a tuple $\T{a} \in T_A$ and a
tuple $\T{b} \in T_B$. We will denote such a configuration by
\conf{\T{a}}{\T{b}}.  The conditions for the duplicator to win the
game with 0 rounds are:

\begin{enumerate}
\item $\forall R \in \dbschema, \forall \mathrm{X} \subseteq
  \{1,\ldots,\ARITY(R) \} : \T{a} \in \pi_{\mathrm{X}}(A(R))
  \Leftrightarrow \T{b} \in \pi_{\mathrm{X}}(B(R))$
\item for every atomic formula (equivalently, for every
  quantifier-free formula) $\theta$ over $\Omega$, $\theta(\T{a})$
  holds iff $\theta(\T{b})$ holds.
\end{enumerate} 

In the game with $m \geq 1$ rounds, the spoiler will be the first one
to make a move. Therefore, he first chooses a database ($A$ or $B$).
Then he picks a tuple in $T_A$ or in $T_B$ respectively. The
duplicator then has to make an ``analogous'' move in the other tuple
space.  When the duplicator can hold this for $m$ times, no matter
what moves the spoiler takes, we say that the duplicator wins the
$m$-round semijoin game on $A$ and $B$.  The ``analogous'' moves for
the duplicator are formally defined as legal answers in the next
definition.
\begin{definition}[legal answer]
  Suppose that at a certain moment in the semijoin game, the
  configuration is \conf{\T{a}}{\T{b}}. If the spoiler takes a tuple
  $\T{c} \in T_A$ in his next move, then the tuples $\T{d} \in T_B$,
  for which the following conditions hold, are legal answers for the
  duplicator:
\begin{enumerate}
\item $\forall R \in \dbschema, \forall \mathrm{X} \subseteq
  \{1,\ldots,\ARITY(R) \}: \T{d} \in \pi_{\mathrm{X}}(B(R))
  \Leftrightarrow \T{c} \in \pi_{\mathrm{X}}(A(R))$
\item for every atomic formula $\theta$ over $\Omega$,
  $\theta(\T{a},\T{c})$ holds iff $\theta(\T{b},\T{d})$ holds.
\end{enumerate}
If the spoiler takes a tuple $\T{d} \in T_B$, the legal answers $\T{c}
\in T_A$ are defined identically.
\end{definition}

In the following, we denote the semijoin game with initial
configuration \conf{\T{a}}{\T{b}} and that consists of $m$ rounds, by
\game{m}{\T{a}}{\T{b}}. 

We first state and prove
\begin{proposition} \label{lem:S_m}
  If the duplicator wins \game{m}{\T{a}}{\T{b}}, then for each
  semijoin expression $E$ with $\leq m$ nested semijoins and
  projections, we have $ \T{a} \in E(A) \Leftrightarrow \T{b} \in
  E(B) $.
\end{proposition}
\begin{proof}
  We prove this by induction on $m$. The base case $m=0$ is clear. Now
  consider the case $m > 0$. Suppose that $\T{a} \in E_1
  \semijoin_{\theta} E_2(A)$ but $\T{b} \not\in E_1 \semijoin_{\theta}
  E_2(B)$. Then $\T{a} \in E_1(A)$ and $\exists \T{c} \in E_2(A):
  \theta(\T{a},\T{c})$, and either (*)~$\T{b} \not\in E_1(B)$ or
  (**)~$\lnot\exists \T{d} \in E_2(B): \theta(\T{b},\T{d})$. In
  situation (*), \T{a} and \T{b} are distinguished by an expression
  with $m-1$ semijoins or projections, so the spoiler has a winning
  strategy; in situation (**), the spoiler has a winning strategy by
  choosing this $\T{c} \in E_2(A)$ with $\theta(\T{a},\T{c})$,
  because each legal answer of the duplicator \T{d} has
  $\theta(\T{b},\T{d})$ and therefore $\T{d} \not\in E_2(B)$. So, the
  spoiler now has a winning strategy in the game
  \game{m-1}{\T{c}}{\T{d}}. In case a projection distinguishes \T{a}
  and \T{b}, a similar winning strategy for the spoiler exists.  In
  case \T{a} and \T{b} are distinguished by an expression that is
  neither a semijoin, nor a projection, there is a simpler expression
  that distinguishes them, so the result follows by structural
  induction.
\end{proof}

We now come to the main theorem of this section. This theorem concerns
the game \game{\infty}{\T{a}}{\T{b}}, which we also abbreviate as
\game{}{\T{a}}{\T{b}}. We say that the duplicator wins
\game{}{\T{a}}{\T{b}} if the spoiler has no winning strategy. This
means that the duplicator can keep on playing forever, choosing legal
answers for every move of the spoiler.

\begin{theorem}
  The duplicator wins \game{}{\T{a}}{\T{b}} if and only if for each
  semijoin expression $E$, we have $\T{a} \in E(A) \Leftrightarrow
  \T{b} \in E(B) $.
\end{theorem}
\begin{proof}
  The `only if' direction of the proof follows directly from
  Proposition~\ref{lem:S_m}, because if the duplicator wins
  \game{}{\T{a}}{\T{b}}, he wins \game{m}{\T{a}}{\T{b}} for every $m
  \geq 0$. So, \T{a} and \T{b} are indistinguishable through all
  semijoin expressions. For the `if' direction, it is sufficient to
  prove that if the duplicator loses, \T{a} and \T{b} are
  distinguishable. We therefore construct, by induction, a semijoin
  expression $E_{\T{a}}^r$ such that (i) $\T{a} \in E_{\T{a}}^r(A) $,
  and (ii) $\T{b} \in E_{\T{a}}^r(B)$ iff the duplicator wins
  \game{r}{\T{a}}{\T{b}}. We define $E_{\T{a}}^0$ as
\[
\sigma_{\theta_{\T{a}}} \big( \bigcap_{R \in \dbschema} \bigcap_{
  \{\mathrm{X} \subseteq Z \mid \T{a} \in \pi_{\mathrm{X}}(A(R)) \}}
\pi_{\mathrm{X}}(R) \big) - \bigcup_{R \in \dbschema} \bigcup_{ \{
  \mathrm{X} \subseteq Z \mid \T{a} \not\in \pi_{\mathrm{X}}(A(R)) \}
} \pi_{\mathrm{X}}(R)
\]
In this expression, $Z$ is a shorthand for $\{ 1,\ldots, \ARITY(R)\}$
and $\theta_{\T{a}}$ is the \emph{atomic type} of \T{a} over $\Omega$,
i.e., the conjunction of all atomic and negated atomic formulas over
$\Omega$ that are true of \T{a}. 
 
We now construct $E_{\T{a}}^r$ in terms of $E_{\T{a}}^{r-1}$:
\[
\bigcap_{\T{c} \in T_A} \big( E_{\T{a}}^0
\semijoin_{\theta_{\T{a},\T{c}}} E_{\T{c}}^{r-1} \big) \cap \big(
E_{\T{a}}^0 - \bigcup_{j=1}^{s} \bigcup_{\theta} \big( E_{\T{a}}^0
\semijoin_{\theta} \bigcap_{\substack{\T{c} \in T_A\\
    \theta(\T{a},\T{c})}} (E_{\T{c}}^{r-1})^{\mathrm{compl}} \big) \big)
\]
In this expression, $\theta_{\T{a},\T{c}}$ is the atomic type of \T{a}
and \T{c} over $\Omega$; $s$ is the maximal arity of a relation in
$\dbschema$; $\theta$ ranges over all atomic $\Omega$-types of two
tuples, one with the arity of \T{a}, and one with arity $j$. The
notation $E^{\mathrm{compl}}$, for an expression of arity $k$, is a
shorthand for
\[
\big( \bigcup_{R \in \dbschema} \bigcup_{ \substack{X \subseteq
    \{1,\ldots,\ARITY(R)\} \\ \#X = k}} \pi_{\mathrm{X}} (R) \big) - E
\]
\end{proof}

\section{Queries inexpressible in the semijoin algebra}
\label{sec:inexpressibility}

Gr{\"a}del~\cite{Graedel99} already showed that transitivity is
not expressible in the guarded fragment. We will now show that
transitivity is still inexpressible in the more powerful semijoin
algebra.

\begin{theorem}
  Transitivity is inexpressible in the semijoin algebra.
\end{theorem}
\begin{proof}
  We will give two databases $A$ and $B$ over the schema $\dbschema$
  containing a single relation $R$, that are indistinguishable by
  semijoin expressions, and with the property that $R$ is transitive
  in $A$ and not in $B$.  These databases are shown graphically in
  Figure~\ref{fig:transitivity}.

  \begin{figure}[htbp]
    \centering \scalebox{0.5}{ \includegraphics{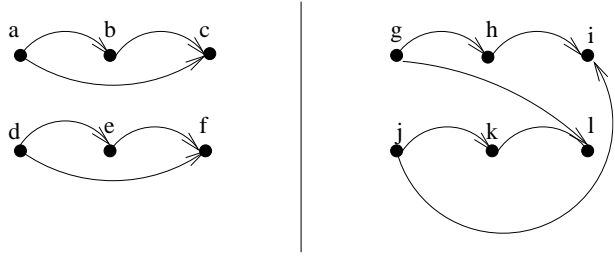}
    }
    \caption{Databases $A$ (left) and $B$ (right) that imply inexpressibility of transitivity in the semijoin algebra}
    \label{fig:transitivity}
  \end{figure}
  
  In this figure the edges represent the relation $R$. A moment's
  inspection reveals that the duplicator has a winning strategy in
  the semijoin game \game{}{\emptytuple}{\emptytuple}.  For the sake
  of completeness we give here the formal strategy. We do this by
  using the following bijections from tuple space $T_A$ to $T_B$.
\[
  \begin{array}[t]{c|c}
    \multicolumn{2}{c}{f: T_A \rightarrow T_B}\\
    \hline
    a \mapsto g\   &  \ ab \mapsto gh\\
    b \mapsto h\   &  \ bc \mapsto hi\\
    c \mapsto i\   &  \ de \mapsto jk\\  
    d \mapsto j\   &  \ ef \mapsto kl\\
    e \mapsto k\   &  \ ac \mapsto gl\\  
    f \mapsto l\   &  \ df \mapsto ji 
  \end{array}
    \hspace{1.0cm}
  \begin{array}[t]{c|c}
    \multicolumn{2}{c}{g: T_A \rightarrow T_B}\\
    \hline
    a \mapsto j\   &  \ ab \mapsto jk\\
    b \mapsto k\   &  \ bc \mapsto kl\\
    c \mapsto l\   &  \ de \mapsto gh\\
    d \mapsto g\   &  \ ef \mapsto hi\\
    e \mapsto h\   &  \ ac \mapsto ji\\
    f \mapsto i\   &  \ df \mapsto gl
  \end{array}
\]
 
When the spoiler makes his first move, the duplicator has a legal
answer by taking the image or pre-image of the spoiler's chosen tuple
under bijection $f$. The duplicator now continues answering each
spoiler move by applying $f$ or $f^{-1}$ to the chosen tuple, until:
 \begin{itemize}
 \item in configuration \conf{ac}{gl} the spoiler chooses $bc$ or
   $kl$, or
 \item in configuration \conf{bc}{hi} the spoiler chooses $ac$ or
   $ji$, or
 \item in configuration \conf{df}{ji} the spoiler chooses $ef$ or
   $hi$, or
 \item in configuration \conf{ef}{kl} the spoiler chooses $df$ or
   $gl$.
 \end{itemize}
 In either case, the duplicator answers with the tuple obtained from
 applying $g$ or $g^{-1}$ to the chosen tuple, and from then, he
 follows strategy function $g$.  Following $g$, he switches back to
 strategy function $f$ whenever:
 \begin{itemize}
 \item in configuration \conf{ac}{ji} the spoiler chooses $bc$ or
   $hi$, or
 \item in configuration \conf{bc}{kl} the spoiler chooses $ac$ or
   $gl$, or
 \item in configuration \conf{df}{gl} the spoiler chooses $ef$ or
   $kl$, or
 \item in configuration \conf{ef}{hi} the spoiler chooses $df$ or
   $ji$.
 \end{itemize}
\end{proof}

Another example of a query inexpressible in the semijoin algebra is
the following: 
\begin{theorem} \label{thm:cart}
  The query $R = \pi_1(R) \times \pi_2(R)$? about a binary relation
  $R$ is inexpressible in the semijoin algebra.
\end{theorem}
\begin{proof}
  In Figure~\ref{fig:cartesian}, two databases $A$ and $B$ are shown
  where $A$ satisfies the query and $B$ does not. The duplicator has a
  winning strategy in the semijoin game
  \game{}{\emptytuple}{\emptytuple}.
  \begin{figure}[htbp]
    \centering \scalebox{0.5}{ \includegraphics{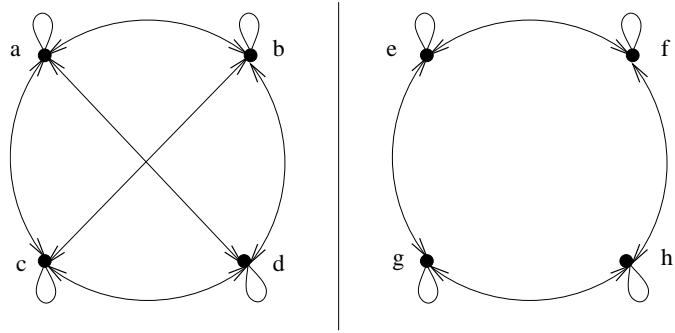} }
    \caption{Databases $A$ (left) and $B$ (right)
      that imply inexpressibility of the query of Theorem~\ref{thm:cart}.}
    \label{fig:cartesian}
  \end{figure}
\end{proof}

\section{Impact of order}
\label{sec:impact-order}

In this section, we investigate the impact of order. On ordered
databases (where $\Omega$ now also contains a total order on the
domain), the query that asks if there are at least $k$ elements in a
unary relation $S$ becomes expressible as $\mathrm{at\_least}(k)$,
which is inductively defined as follows:
\[ \left\{ \begin{array}{rcl}
    \mathrm{at\_least}(1) & := & S\\
    \mathrm{at\_least}(k) & := & S \ltimes_{x_1<y_1}
    \big(\mathrm{at\_least}(k-1)\big)
          \end{array} 
        \right.
\]
Note that this query is independent of the chosen order. This
parallels the situation in first-order logic, where there also exists
an order-invariant query that is expressible with but inexpressible
without order (\cite[Exercise 17.27]{AHV95} and \cite[Proposition
2.5.6]{EF99}).

Transitivity remains inexpressible in the semijoin algebra even on
ordered databases. Consider the following databases $A$ and $B$ over a
single binary relation $R$: $A(R)$ is the union of $X$, $Y$ and $Z$,
where $X = \{1,\ldots,m \} \times \{2m+1\}$, $Y = \{ 2m+1 \} \times
\{m+1,\ldots, 2m \}$, and $Z = \{1,\ldots,m \} \times \{m+1,\ldots,2m
\}$; $B(R) = A(R) - \{(\frac{m+1}{2},m+\frac{m+1}{2}) \}$. Clearly,
$R$ is transitive in $A$, but not in $B$. We have shown
elsewhere~\cite{LTV04} that when $m=2n+1$, the duplicator has a
winning strategy in the $n$-round semijoin game
\game{n}{\emptytuple}{\emptytuple}. By Proposition~\ref{lem:S_m},
transitivity is not expressible in SA with order.

\end{document}